\documentstyle[11pt,aaspp4]{article}

\load{\scriptsize}{\sc}

\input{psfig.tex}

\slugcomment{To appear in The Astrophysical Journal Letters.}

\begin{document}


\title{BUMPY SPIN-DOWN OF ANOMALOUS X-RAY PULSARS: \\
 THE LINK WITH MAGNETARS}

\author{A. Melatos\altaffilmark{1,2}}
\affil{Department of Astronomy, 601 Campbell Hall,
 University of California at Berkeley, \\
 Berkeley CA 94720 USA}

\altaffiltext{1}{Miller Fellow}
\altaffiltext{2}{e-mail: melatos@astraea.berkeley.edu}

\begin{abstract}
The two anomalous X-ray pulsars (AXPs) with
well-sampled timing histories,
1E 1048.1$-$5937 and 1E 2259$+$586, 
are known to spin down irregularly,
with `bumps' superimposed on an overall linear trend.
Here we show that if AXPs are non-accreting magnetars,
i.e.\ isolated neutron stars with surface magnetic fields
$B_0 \gtrsim 10^{10}\,{\rm T}$,
then they spin down electromagnetically
in exactly the manner observed,
due to an effect called `radiative precession'.
Internal hydromagnetic stresses deform the star,
creating a fractional
difference $\epsilon=(I_3-I_1)/I_1\sim 10^{-8}$
between the principal moments of inertia $I_1$ and $I_3$;
the resulting Eulerian precession
couples to an oscillating component of the electromagnetic 
torque associated with the near-zone radiation fields,
and the star executes
an anharmonic wobble with period
$\tau_{\rm pr} \sim
 2\pi / \epsilon\Omega(t) \sim 
 10\,{\rm yr}$,
where $\Omega(t)$ is the rotation frequency
as a function of time $t$.
We solve Euler's equations for a biaxial magnet 
rotating {\em in vacuo};
show that the computed $\Omega(t)$ matches the measured
timing histories of 1E 1048.1$-$5937 and 1E 2259$+$586;
predict $\Omega(t)$ for the next 20 years for
both objects;
predict a statistical relation between
$\langle d\Omega/dt \rangle$ and $\tau_{\rm pr}$,
to be tested as the population of known AXPs grows;
and hypothesize
that radiative precession will be observed
in future X-ray timing of soft gamma-ray repeaters (SGRs).
\end{abstract}

\clearpage


\section{INTRODUCTION
 \label{sec:axp1}}
Anomalous X-ray pulsars (AXPs)
are a sub-class of X-ray pulsars distinguished by
pulse periods lying in a narrow range between
$6\,{\rm s}$ and $12\,{\rm s}$,
low X-ray luminosities,
soft X-ray spectra,
no detected optical counterparts,
no detected
orbital Doppler shifts of pulse arrival times,
and associations with shell-type supernova remnants
(\markcite{mer95b}Mereghetti \& Stella 1995;
\markcite{van95}van Paradijs, Taam \& van den Heuvel 1995).
The nature of AXPs is uncertain; possibilities include
an accreting neutron star in a binary, with a low-mass
white dwarf or He-burning star as a companion
(\markcite{mer95b}Mereghetti \& Stella 1995;
\markcite{bay96}Baykal \& Swank 1996;
\markcite{mer98}Mereghetti, Israel \& Stella 1998),
an isolated neutron star accreting from a residual disk
following a phase of common-envelope evolution
(\markcite{van95}van Paradijs et al.\ 1995;
\markcite{gho97}Ghosh, Angelini \& White 1997),
a non-accreting massive white dwarf
(\markcite{mor88}Morini et al.\ 1988;
\markcite{pac90}Paczy\'{n}ski 1990;
\markcite{uso94}Usov 1994),
and a non-accreting magnetar
(\markcite{tho96}Thompson \& Duncan 1996;
\markcite{hey98}Heyl \& Hernquist 1998;
\markcite{kou98}Kouveliotou et al.\ 1998).

The two AXPs with
well-sampled timing histories,
1E 1048.1$-$5937 and 1E 2259$+$586, 
are observed to spin down irregularly:
the rotation frequency $\Omega$ of each object decreases 
linearly with time $t$ on average, but there are `bumps' 
superimposed on this average trend
during which $\dot{\Omega}=d\Omega / dt < 0$ 
fluctuates by a factor of between two and five 
every five to ten years 
(\markcite{mer95a}Mereghetti 1995;
\markcite{bay96}Baykal \& Swank 1996;
\markcite{bay98}Baykal et al.\ 1998;
\markcite{oos98}Oosterbroek et al.\ 1998;
and references therein).
In accreting-star models of AXPs, 
the average spin-down trend is
attributed to the accretion torque acting on a 
near-equilibrium rotator, 
and the bumps are the result of white torque
noise (e.g.\ due to disk inhomogeneities)
as in ordinary binary X-ray pulsars
(\markcite{bay96}Baykal \& Swank 1996).
In isolated-star models, the spin-down trend 
is attributed to magnetic-dipole braking,
and the bumps are analogous to glitches
observed in rotation-powered pulsars like Vela
(\markcite{uso94}Usov 1994;
\markcite{hey98}Heyl \& Hernquist 1998).

In this Letter, we present unequivocal new evidence 
that AXPs are non-accreting magnetars.
In \S\ref{sec:axp2}, 
it is shown that as a magnetar spins down it wobbles
anharmonically, with a period of five to ten years, 
due to an effect called `radiative precession'.
The spin-down signature of the wobble,
calculated theoretically in \S\ref{sec:axp3},
matches closely the bumpy timing
histories of 1E 1048.1$-$5937 and 1E 2259$+$586.
The theory is used to predict $\Omega(t)$
over the next 20 years for both objects
and yields a statistical relation
between bump recurrence time
and average spin-down rate for the AXP population 
as a whole.
Implications for the internal and magnetospheric 
structures of magnetars, including AXPs and 
soft gamma-ray repeaters (SGRs),
are explored in \S\ref{sec:axp4}.

\section{RADIATIVE PRECESSION
 \label{sec:axp2}}
A magnetar is a triaxial body in general.
Hydromagnetic stresses arising from non-radial gradients 
of the superstrong internal magnetic field,
e.g.\ between the magnetic poles and equator if the field 
is dipolar,
deform the star, inducing matter-density perturbations 
$\delta\rho\sim B_{\rm in}^2/\mu_0 c_{\rm s}^2$
and hence a fractional difference
$\epsilon\sim \delta\rho R^5 / I_1
 \approx 2\times 10^{-9} (B_{\rm in}/10^{10}\,{\rm T})^2$
between the principal moments of inertia
(\markcite{gol70}Goldreich 1970;
\markcite{dec80}de Campli 1980;
\markcite{mel98}Melatos 1998).
Here,
$B_{\rm in}$ is the characteristic magnitude of the
internal magnetic field, $c_{\rm s}$ is the isothermal
sound speed ($c_{\rm s}\approx 3^{-1/2}c$),
and $R$ is the stellar radius;
one has $B_{\rm in}\approx B_0$ if the internal 
field is confined to the stellar crust
and $B_{\rm in}\gtrsim B_0$ if it
is generated deep inside the star, 
e.g.\ in a convective dynamo
(\markcite{tho93}Thompson \& Duncan 1993).
In a rotation-powered pulsar with
$B_0\lesssim 10^9\,{\rm T}$,
the hydromagnetic deformation is comparable to the
elastic deformation arising from shear stresses in the
crystalline stellar crust,
and the principal axes of inertia are oriented
arbitrarily with respect to the axis ${\bf m}$
of the external magnetic dipole
(\markcite{gol70}Goldreich 1970;
\markcite{dec80}de Campli 1980;
\markcite{mel98}Melatos 1998).
In a magnetar, however, the hydromagnetic deformation is
much larger, and ${\bf m}$ is approximately parallel to
one of the principal axes (${\bf e}_3$, say);
the alignment is not exact
due to the complicated structure
of the internal field near its generation site,
cf.\ the non-axisymmetric, magnetically modified Taylor
columns seen in simulations of the geodynamo
(\markcite{gla96}Glatzmaier \& Roberts 1996).
Provided that the rotation axis ${\bf \Omega}$ is
not parallel to ${\bf m}\approx {\bf e}_3$,
as is usually the case for rotation-powered pulsars,
the star precesses about ${\bf e}_3$ 
with period 
$\tau_{\rm pr}
\sim 2\pi/\epsilon\Omega
\approx 85(B_{\rm in}/10^{10}\,{\rm T})^{-2}
 (\Omega/1\,{\rm rad\,s^{-1}})^{-1}\,{\rm yr}$.

The Eulerian precession is not free. It couples
to a component of the vacuum magnetic-dipole
torque associated with the asymmetric inertia 
of the near-zone radiation fields 
outside the magnetar;
the electromagnetic energy density (and hence inertia)
of the near-zone radiation fields is greater at the 
magnetic poles than at the equator by an amount
$\sim B_0^2/\mu_0$,
resulting in an oscillatory, precession-inducing
torque 
(\markcite{gol70}Goldreich 1970;
\markcite{goo85}Good \& Ng 1985;
\markcite{mel98}Melatos 1998).
The near-field torque exceeds the familiar
braking torque ($\propto\Omega^3$)
by a factor $c/\Omega R\gg 1$ and acts on a
commensurately shorter time-scale,
$\tau_{\rm nf}\sim\tau_0 \Omega R/c
 \approx
 6 (B_0/10^{10}\,{\rm T})^{-2}
 (\Omega/1\,{\rm rad\,s^{-1}})^{-1}\,{\rm yr}$,
where 
$\tau_0=\mu_0 c^3 I_1/2\pi B_0^2 R^6 \Omega^2
 \approx
 2\times 10^5 (B_0/10^{10}\,{\rm T})^{-2}
 (\Omega/1\,{\rm rad\,s^{-1}})^{-2}{\rm yr}$
is the characteristic electromagnetic braking time.
Since the near-field torque is directed along
${\bf \Omega}\times {\bf m}$, it does not
change $\Omega$ for a spherical star.
Nor does it change $\Omega$ for an aspherical star,
provided $\epsilon$ is large enough
to give $\tau_{\rm pr}\ll\tau_{\rm nf}$,
so that the near-field torque averages to zero over
one precession period.
When the dominant deformation is hydromagnetic,
however,
one finds 
$\tau_{\rm pr}/\tau_{\rm nf}
 \approx
 14 (B_0/B_{\rm in})^2$,
close to unity (i.e.\ strong coupling)
provided $B_{\rm in}$
is moderately larger than $B_0$ as expected
(\markcite{tho93}Thompson \& Duncan 1993).
Under these circumstances, 
the star executes an anharmonic wobble,
called radiative precession,
with period $\tau_{\rm pr}$ ($\approx\tau_{nf}$),
and the near-field torque changes $\Omega$ on
the precession time-scale
(\markcite{mel98}Melatos 1998).

\section{THEORY OF BUMPY SPIN-DOWN
 \label{sec:axp3}}

\subsection{Solution of Euler's Equations of Motion:
 Past and Future $\Omega (t)$
 \label{sec:axp3a}}
We now show that the timing signature of radiative 
precession matches the observed bumpy spin-down of AXPs
by solving numerically Euler's equations of motion for a 
biaxial, dipole magnet rotating {\em in vacuo},
\begin{eqnarray}
\dot{\Omega}_1
 & = &
 -\epsilon\Omega_2\Omega_3
 + \Omega_0^{-2}\tau_0^{-1}\cos\chi
 [ a\Omega^2(-\Omega_1\cos\chi+\Omega_3\sin\chi) +
   b\Omega_2(\Omega_1\sin\chi+\Omega_3\cos\chi) ],
 \label{eq:axp1}
 \\
\dot{\Omega}_2
 & = &
 \epsilon\Omega_1\Omega_3
 + \Omega_0^{-2}\tau_0^{-1}
 [ -a\Omega^2\Omega_2 +
   b(-\Omega_1\cos\chi+\Omega_3\sin\chi)
    (\Omega_1\sin\chi+\Omega_3\cos\chi) ],
 \label{eq:axp2}
 \\
\dot{\Omega}_3
 & = &
 - \Omega_0^{-2}\tau_0^{-1}\sin\chi
 [ a\Omega^2(-\Omega_1\cos\chi+\Omega_3\sin\chi) +
   b\Omega_2(\Omega_1\sin\chi+\Omega_3\cos\chi) ].
 \label{eq:axp3}
\end{eqnarray}
In (\ref{eq:axp1})--(\ref{eq:axp3}),
subscripts denote components along the 
principal axes of inertia,
$\chi$ is the angle between ${\bf m}$ and ${\bf e}_3$,
and we have $a=0.33$, $b=0.094c/\Omega_0 R$,
and $\Omega_0=\Omega(t=t_0)$, 
where $t_0$ is an arbitrary origin;
for derivations of the equations, see 
\markcite{gol70}Goldreich (1970) and
\markcite{mel98}Melatos (1998).
Terms in (\ref{eq:axp1})--(\ref{eq:axp3})
proportional to $\epsilon$ give rise to Eulerian precession,
terms proportional to $b\tau_0^{-1}$ are associated with
the near-field torque,
and terms proportional to $a\tau_0^{-1}$ cause
secular braking.
A biaxial, dipole magnet rotating {\em in vacuo}
is an idealized model of a hydromagnetically 
deformed magnetar.
In reality, such a body is triaxial 
(if it is indeed rigid),
has high-order and/or off-centered multipoles contributing 
to the near-zone magnetic field,
and is surrounded by a plasma magnetosphere.
The values of $a$ and $b$ reflect,
in a coarse way,
the magnetization state of the stellar interior
and the distribution of magnetospheric currents
(\markcite{mel98}Melatos 1998).

\begin{figure}
\plotone{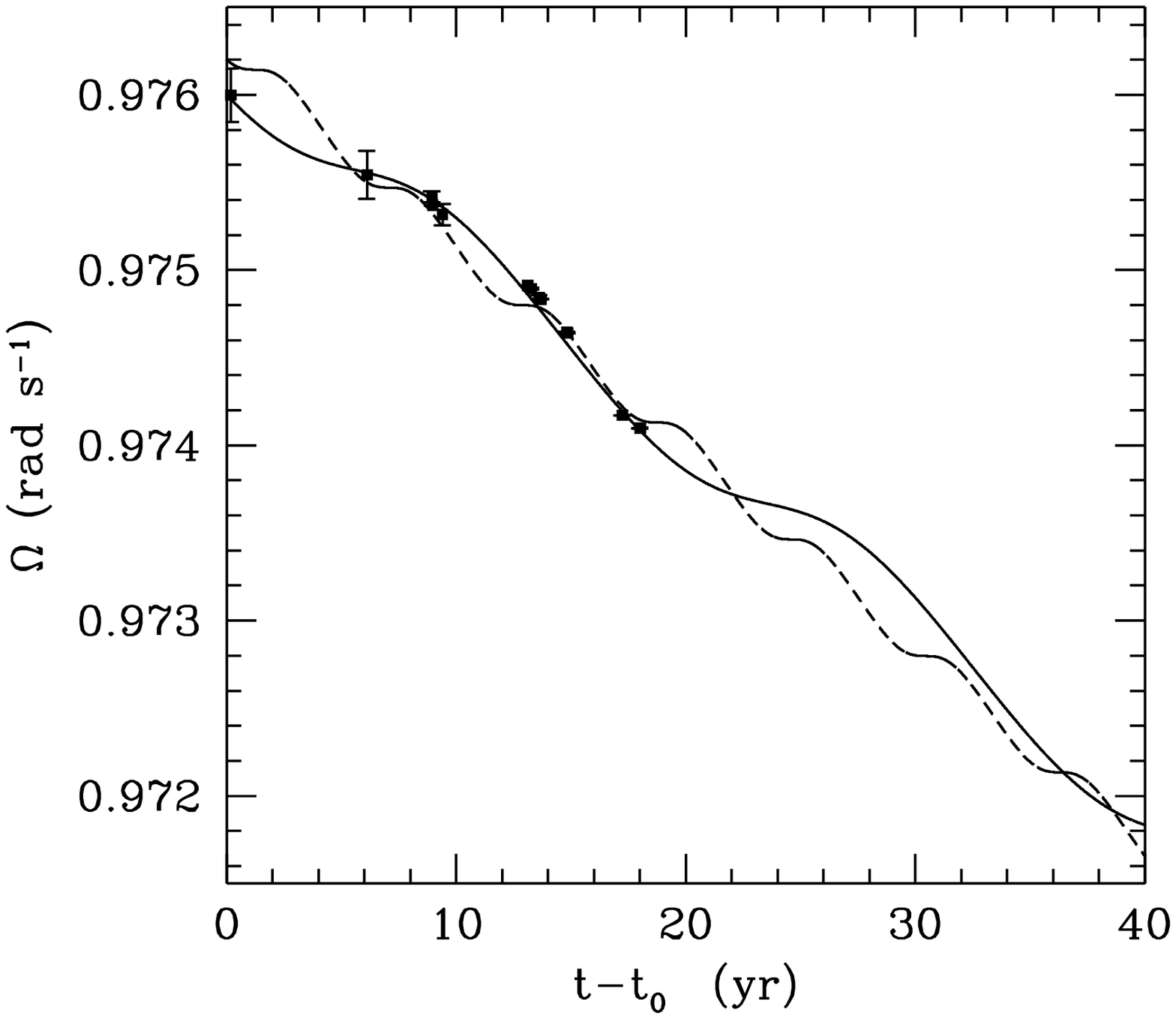}
\caption{
Rotation frequency $\Omega$ versus time $t$ for the AXP
1E 1048.1$-$5937, 
with $t_0={\rm JD}\,\,2444000$.
The squares and accompanying $1\sigma$
error bars are X-ray timing measurements
made over the last 20 years by the satellites 
{\em Einstein}, {\em EXOSAT}, {\em Ginga}, {\em ROSAT},
{\em ASCA}, {\em RXTE}, and {\em BeppoSAX} 
(Oosterbroek et al.\ 1998, and references therein).
The solid curve is the solution to Euler's equations
of motion (\ref{eq:axp1})--(\ref{eq:axp3}) for
$\Omega_0=0.976\,{\rm rad\,s^{-1}}$,
$\Omega_0\tau_0=3.8\times 10^{10}$,
$\epsilon=6.4\times 10^{-8}$, and
$\chi=3.5^\circ$,
with initial conditions
$\Omega_{1,0}=0.476\Omega_0$ and
$\Omega_{2,0}=-0.568\Omega_0$.
The broken curve is the solution for
$\Omega_0=0.9762\,{\rm rad\,s^{-1}}$,
$\Omega_0\tau_0=3.5\times 10^{10}$,
$\epsilon=6.3\times 10^{-8}$, and
$\chi=18^\circ$,
with the same initial conditions.
The initial conditions determine the initial phase of
the oscillation and are otherwise insignificant.
}
\label{fig:axp1}
\end{figure}

\begin{figure}
\plotone{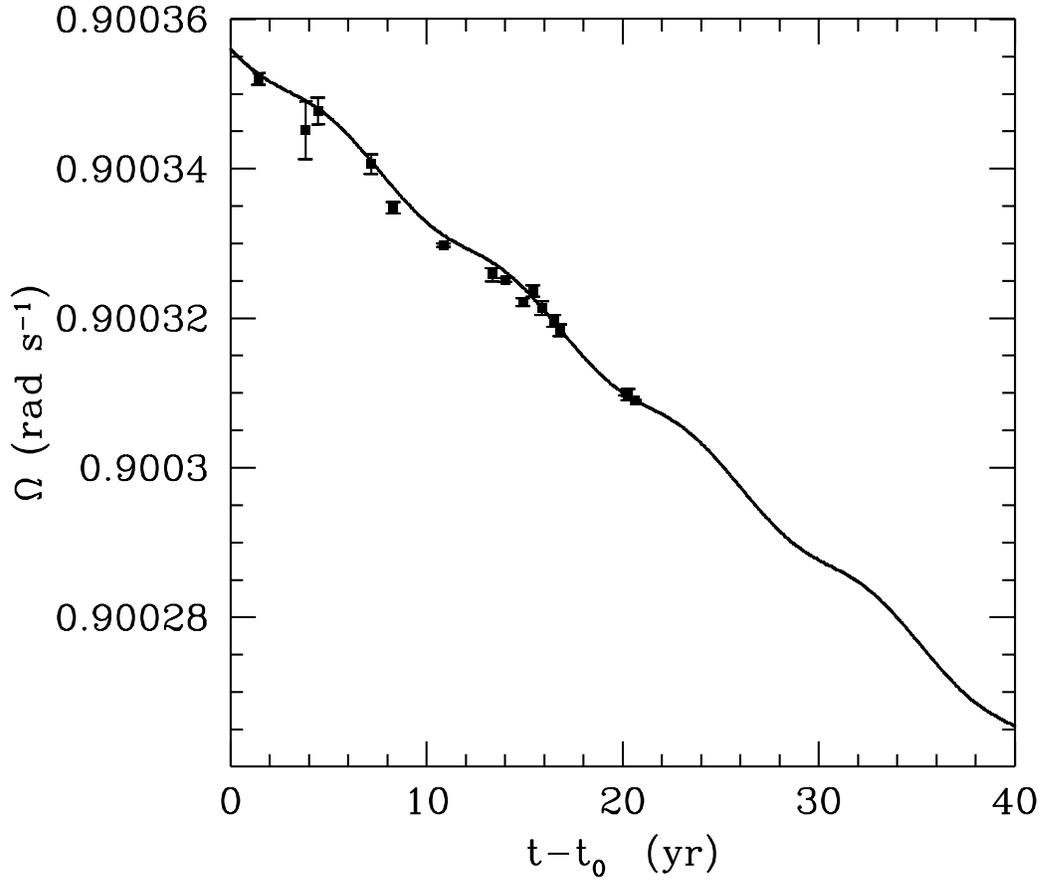}
\caption{
Rotation frequency $\Omega$ versus time $t$ for the AXP
1E 2259$+$586, 
with $t_0={\rm JD}\,\,2443000$.
The squares and $1\sigma$ error bars
are X-ray timing data from the
satellites cited in Figure \ref{fig:axp1} as well as from
{\em HEAO 1} and {\em Tenma} 
(Baykal et al.\ 1998, and references therein).
The solid curve is the solution to Euler's equations
of motion (\ref{eq:axp1})--(\ref{eq:axp3}) for
$\Omega_0=0.900356\,{\rm rad\,s^{-1}}$,
$\Omega_0\tau_0=2.35\times 10^{12}$,
$\epsilon=3.4\times 10^{-8}$, and
$\chi=13^\circ$,
with initial conditions
$\Omega_{2,0}=-0.668\Omega_0$ and
$\Omega_{3,0}=0.658\Omega_0$.
}
\label{fig:axp2}
\end{figure}

In Figures \ref{fig:axp1} and \ref{fig:axp2},
we plot the computed $\Omega(t)$ from
(\ref{eq:axp1})--(\ref{eq:axp3})
on top of the X-ray timing histories of 
1E 1048.1$-$5937 and 1E 2259$+$586 respectively,
extending the theoretical curves 20 years beyond the 
present as a testable prediction.
The timing history of 1E 2259$+$586 is sufficiently
well-sampled that only one good fit is possible.
In the case of 1E 1048.1$-$5937, the solid curve is the
favored fit, with $\tau_{\rm pr}\approx 18\,{\rm yr}$, 
but an alternative fit, with similar average slope and 
$\tau_{\rm pr}\approx 6\,{\rm yr}$, 
is also acceptable.
Multiple good fits are hard to find.

Euler's equations of motion contain just three unknown
parameters: $\epsilon$, $\tau_0$, and $\chi$.
It is significant that the theory agrees with
the observations as well as it does
despite its idealized nature ---
particularly as only two of the three parameters
are truly free, since one needs strong torque
coupling
($\tau_{\rm pr}\sim\tau_{\rm nf}$,
or equivalently
$\epsilon\Omega_0\tau_0\sim c/\Omega_0 R$)
in order to get any bumps at all.
Moreover, the best agreement with observations is
achieved for parameter values that are consistent 
with general physical considerations.
If AXPs are hydromagnetically deformed magnetars
with
$B_{\rm in}\gtrsim B_0\gtrsim 3\times 10^{10}\,{\rm T}$,
one expects
$\epsilon\sim 10^{-8}$,
$\Omega_0\tau_0\sim 10^{11}$, 
and $\chi$ relatively small
(cf.\ $\chi\approx 11^\circ$ for the Earth),
as discussed above.

The theoretical curves do not pass exactly through every 
available data point, and more departures are
expected in the future,
e.g.\ the slope between the last two data points
in Figure \ref{fig:axp1} is $\approx 0.6$ times
the solid-curve theoretical slope at that epoch.
Indeed, a formal estimate of the
chi-square for the solid curve in Figure \ref{fig:axp1}, 
taking published timing uncertainties at face value, 
implies that the fit is poor:
one finds a chi-square of $\approx 8\times 10^3$ with 10 degrees
of freedom, notably inferior 
to multiple-glitch models, for example
(\markcite{hey98}Heyl \& Hernquist 1998).
This is because a biaxial, dipole magnet is an
over-idealized model of an AXP, as noted above;
the chi-square likelihood improves dramatically for
a more realistic model with just two extra parameters,
e.g.\ a non-dipolar near-zone magnetic field
and a triaxial ellipsoid of inertia.
However, our aim in this paper is not to model
$\Omega(t)$ in detail, but to account for key
gross features of the data ---
the average spin-down rate,
bump recurrence time, and bump amplitude ---
with a simple physical theory.
In this regard, the agreement with observations
is good.

One data point in the timing history of 1E 2259$+$586, 
at $t-t_0=15.4\,{\rm yr}$,
represents a spin-up event 
lasting at most $0.8\,{\rm yr}$
at the $1\sigma$ level of uncertainty
(\markcite{bay96}Baykal \& Swank 1996).
Spin-up cannot be explained
by radiative precession
because (i) $\dot{\Omega}$ is always negative,
even while
$|\dot{\Omega}|$ decreases by up to a factor of five
during a bump,
and (ii)
there is no natural way to explain a $0.8\,{\rm yr}$
time-scale.
If taken at face value, the spin-up event must 
be a different phenomenon,
e.g.\ a discontinuous change of internal structure
analogous to the glitches of rotation-powered 
pulsars like Vela
(\markcite{uso94}Usov 1994;
\markcite{hey98}Heyl \& Hernquist 1998).

\subsection{Predicted Population Statistics
 \label{sec:axp3b}}
Numerical studies show that
the bump recurrence time and average spin-down rate 
satisfy
$\tau_{\rm pr}\propto 
 B_{\rm in}^{-2}\Omega^{-1} f(\chi)$
and
$\langle\dot{\Omega}\rangle\propto
 B_0^2\Omega^3 g(\chi)$,
with $1\leq f(\chi), \, g(\chi) \leq 10$.
In other words,
the greater the magnetic field of an AXP,
the shorter is its precession period and the greater 
is its spin-down rate, with
\begin{equation}
 \langle\dot{\Omega}\rangle \approx
 -2\times 10^{-4} (B_0/B_{\rm in})^2
 (\Omega/1\,{\rm rad\,s^{-1}})^2
 (\tau_{\rm pr}/ 1\,{\rm yr})^{-1}
 \,{\rm rad\,s^{-1}\,yr^{-1}}.
 \label{eq:axp4}
\end{equation}
In addition,
one finds that $\dot{\Omega}$ increases
from
$\dot{\Omega}\approx \langle\dot{\Omega}\rangle$ to
$\dot{\Omega}\approx 0$ 
over a time
$\approx 0.25\tau_{\rm pr}$
during the course of a bump,
yielding a bump amplitude
\begin{equation}
 \Delta\Omega_{\rm pr}/\Omega \approx
 5\times 10^{-5} (B_0/B_{\rm in})^2
 (\Omega/1\,{\rm rad\,s^{-1}})~;
 \label{eq:axp5}
\end{equation}
this number is similar for all AXPs.
Both the relations (\ref{eq:axp4}) and (\ref{eq:axp5})
will be subject to observational testing
as the population of AXPs with measured timing histories
swells over time. 
Note that they are statistical relations,
with scatter expected about an overall trend.
The detailed structure of the magnetic field
inside an AXP --- which does not enter into the
idealized theory presented here, except through $\chi$ ---
is likely to differ from object to object,
affecting
$\tau_{\rm pr}$, $\langle\dot{\Omega}\rangle$,
and $\Delta\Omega_{\rm pr}$ significantly,
as the broad ranges of $f(\chi)$ and $g(\chi)$ attest.

\section{DISCUSSION 
 \label{sec:axp4}}
Why is radiative precession not observed in
rotation-powered pulsars with $B_0\lesssim 10^9\,{\rm T}$?
These objects are deformed hydromagnetically like
magnetars, with an added elastic deformation,
and they spin down electromagnetically.
Young pulsars like the Crab, with
$\tau_0\approx 10^3\,{\rm yr}$ and 
$\Omega R/c\approx 10^{-2}$,
ought to precess with period 
$\tau_{\rm pr}\approx 10\,{\rm yr}$,
yet there is no clean evidence of bumpy spin-down
in radio timing data, nor of concomitant
changes in pulse profile (e.g.\ relative height and
separation of conal components) and polarization
characteristics (e.g.\ position-angle swing).
\markcite{lyn88}Lyne, Pritchard \& Smith (1988) reported
a quasi-sinusoidal variation
in Crab timing residuals with a period of $\approx 20$
months but judged it likely to be an artifact
of an overlooked glitch.
(See also \markcite{mel98}Melatos 1998.)
The only reliable detection of pulsar precession
to date has been the general-relativistic geodetic
precession of the binary pulsar PSR B1913$+$16
(\markcite{wei89}Weisberg, Romani \& Taylor 1989;
\markcite{kra98}Kramer 1998).

One possible explanation is that radiative precession 
is damped in pulsars with $B_0\lesssim 10^9\,{\rm T}$.
Frictional dissipation inside the star, due to 
time-dependent elastic strains in the crust 
and/or imperfect crust-core coupling, is thought to
occur on a time-scale $\lesssim 1\,{\rm yr}$ 
(\markcite{lin93}Link, Epstein \& Baym 1993),
rapidly aligning ${\bf \Omega}$ with ${\bf e}_3$.
(In the case of the Earth, dissipation restricts
${\bf \Omega}$ to fluctuating within just
$1''$ of ${\bf e}_3$ under the
action of solar and lunar tides;
see Bur\v{s}a \& P\v{e}\v{c} 1993).
In a magnetar, where the magnetic energy exceeds
the mechanical energy of rotation,
the stiffening effect of the superstrong magnetic field 
may hinder the elastic strains and/or sheared fluid flows 
responsible for internal damping.
A second possible explanation is that conduction
currents in the magnetosphere of a pulsar
with $B_0\lesssim 10^9\,{\rm T}$, 
where electron-positron pair production is plentiful,
modify the electromagnetic torque in such a way as
to suppress the precession-inducing near-field
component.
In a magnetar, where it is thought that pair production
is quenched, e.g.\ by positronium formation
(\markcite{uso96}Usov \& Melrose 1996;
J.\ Heyl 1999, private communication)
or photon splitting
(\markcite{bar98}Baring \& Harding 1998),
the vacuum magnetic-dipole torque,
with its unmodified near-field component,
is a closer approximation to reality.
Both explanations imply that radiative precession will
not be suppressed in SGRs, where one has
$B_0\sim 10^{11}\,{\rm T}$ 
(\markcite{kou98}Kouveliotou et al.\ 1998).

What does radiative precession teach us about strongly
magnetized neutron stars themselves? 
Firstly, the close agreement
between theory and observation implies that AXPs are indeed
non-accreting magnetars, and that accretion is not needed
to explain fluctuations in $\dot{\Omega}$, contrary to
claims in the literature
(e.g.\ \markcite{mer95a}Mereghetti 1995).
Secondly, if SGRs are magnetars too,
as indicated by recent X-ray timing data
(\markcite{kou98}Kouveliotou et al.\ 1998),
they ought to exhibit bumpy spin-down just like AXPs,
perhaps punctuated by brief, glitch-like spin-up events
during the gamma-ray bursts themselves.
This constitutes a direct test of the magnetar model for SGRs.
Thirdly, the fact that bumpy spin-down is seen in AXPs
implies that $B_{\rm in}$ is at most a few times $B_0$
to ensure strong coupling, 
i.e.\ $\tau_{\rm pr}\sim \tau_{\rm nf}$,
as discussed above.
This is indirect evidence that the magnetic field of
these objects is generated relatively near the stellar
surface.
Fourthly, radiative precession 
places an upper limit on the fluid viscosity $\eta$
of a newly born magnetar; if $\eta$ is too high,
${\bf \Omega}$ aligns with ${\bf e}_3\approx{\bf m}$ 
before the stellar crust
crystallizes, and there is no precession
(\markcite{mel98}Melatos 1998).
The upper limit obtained in this way --- 
that the viscous damping time
is greater than the crust crystallization time of 
$\sim 1\,{\rm yr}$ ---
validates certain semi-quantitative calculations of $\eta$ for
pulsars with $B_0\lesssim 10^9\,{\rm T}$ by 
\markcite{cut87}Cutler \& Lindblom (1987).
We remark in closing that the predicted hydromagnetic
deformation ($\epsilon\sim 10^{-8}$)
constitutes a misaligned mass quadrupole
which generates gravitational radiation.
However, the radiation is too weak to be detected by 
planned gravitational-wave interferometers like LIGO
and VIRGO, because AXPs are slow rotators and the
gravitational-wave amplitude is a strongly rising
function of $\Omega$.


\acknowledgments
I am indebted to Lars Bildsten for bringing the existence 
of spin-down irregularities in AXPs to my attention,
and for subsequent discussions. 
I also thank Jon Arons, Peter Goldreich, and the referee,
Jeremy Heyl, for comments.
This work was supported by NASA Grant NAG5--3073,
NSF Grant AST--95--28271,
and by the Miller Institute for Basic Research in Science
through a Miller Fellowship.




\clearpage

\end{document}